\documentclass[aps,amsmath,amssymb,pra,groupedaddress,superscriptaddress,floatfix,preprint]{revtex4-2}

\usepackage[T2A]{fontenc}
\usepackage[utf8]{inputenc}
\usepackage[english]{babel}

\usepackage{amsmath,amsfonts,amssymb,amsthm,mathtools}
\usepackage{graphicx}
\usepackage{nicefrac}
\usepackage{epsf}
\usepackage{bm}
\usepackage{braket}
\usepackage{dcolumn}
\usepackage{multirow}
\usepackage{booktabs}
\usepackage{hyperref}
\usepackage{float}
\usepackage{bbold}
\usepackage{colortbl}
\usepackage{xcolor}
\usepackage{color}
\usepackage[normalem]{ulem}
\usepackage{longtable}
\usepackage{siunitx}
\usepackage{etoolbox}
\usepackage{threeparttable}
\usepackage{svg}
\usepackage{soul}
\usepackage{tabularx}

\mathtoolsset{showonlyrefs=false}

\newcolumntype{C}{>{\centering\arraybackslash}X}
\newcolumntype{R}{>{\raggedleft\arraybackslash}X}

\begin{document}
\title{Finite-Basis-Set Approach to the Two-Center Heteronuclear \\ Dirac Problem}

\author{A.~A.~Kotov}
\affiliation{Department of Physics, St. Petersburg State University, 199034 St. Petersburg, Russia}

\author{D.~A.~Glazov}
\affiliation{Department of Physics, St. Petersburg State University, 199034 St. Petersburg, Russia}

\author{A.~V.~Malyshev}
\affiliation{Department of Physics, St. Petersburg State University, 199034 St. Petersburg, Russia}

\author{V.~M.~Shabaev}
\affiliation{Department of Physics, St. Petersburg State University, 199034 St. Petersburg, Russia}

\author{G.~Plunien}
\affiliation{Institut f\"ur Theoretische Physik, \\ Technische Universit\"at Dresden, D-01062 Dresden, Germany}

\begin{abstract}
The rigorous two-center approach based on the dual-kinetically balanced finite-basis-set expansion is applied to one-electron, heteronuclear diatomic Bi-Au, U-Pb, and Cf-U quasimolecules. The obtained $1\sigma$ ground-state energies are compared with previous calculations, when possible. Upon analysis of three different placements of the coordinate system's origin in the monopole approximation of the two-center potential: (1) in the middle, between the nuclei, (2) in the center of the heavy nucleus, and (3) in the center of the light nucleus, a substantial difference between the results is found. The leading contributions of one-electron quantum electrodynamics (self-energy and vacuum polarization) are evaluated within the monopole approximation as well.
\end{abstract}


\maketitle
\newpage

\section{Introduction}
Heavy ion and atom encounters lead to the short-time formation of diatomic quasimolecules.
Presently, collisions of highly charged ions with neutral atoms, e.g., Xe$^{54+}$--Xe, are available for experimental investigation at the GSI Helmholtz Center for Heavy Ion Research ~\cite{Verma:2006_1, Verma:2006_2, Hagmann:2011, Zhu:2022, Hillenbrand:2022}.
The upcoming experiments at NICA~\cite{Ter-Akopian:2015}, HIAF~\cite{Ma:2017}, and GSI/FAIR~\cite{Gumberidze:2009} will allow for observation of heavy few-electron systems' collisions up to bare nuclei, such as U$^{92+}$--U$^{92+}$.

The theoretical prediction of quasimolecular spectra plays an important role in both the study of critical phenomena in bound-state quantum electrodynamics (BS-QED) and the interpretation of experimental data.
A number of theoretical approaches have been developed to investigate the relativistic dynamics of these systems; see Refs.~\cite{Soff:1979, Becker:1986, Eichler:1990, Rumrich:1993, Ionescu:1999, Tupitsyn:2010, Tupitsyn:2012, Maltsev:2019, Popov:2020, Voskresensky:2021} and references therein.
Within the Born--Oppenheimer approximation, the Dirac problem of quasimolecular systems has also been investigated in a number of works~\cite{Muller:1973, Rafelski:1976_1, Rafelski:1976_2, Lisin:1977, Soff:1979, Lisin:1980, Yang:1991, Parpia:1995, Deineka:1998, Matveev:2000, Kullie:2001, Ishikawa:2008, Artemyev:2010, Tupitsyn:2010, Ishikawa:2012, Tupitsyn:2014, Mironova:2015, Artemyev:2015, Artemyev:2022}.
While most of these approaches rely on the partial-wave expansion of the two-center potential, several works have investigated the usage of the Cassini coordinate system ~\cite{Artemyev:2010} and the Dirac--Fock--Sturm method ~\cite{Tupitsyn:2014, Mironova:2015}.

Previously, we considered the one- and two-electron homonuclear quasimolecules of xenon, lead, and uranium in both the rigorous two-center approach and the monopole approximation within the dual-kinetically balanced finite-basis-set approach~\cite{Kotov:2020, Kotov:2021}.
We showed that the obtained solution is in good agreement with other independent calculations of the energy spectra.
In Ref.~\cite{Kotov:2021}, it was shown that an analysis of different placements of the coordinate system's origin~(c.s.o.) could provide an estimation of the non-monopole correction to contributions that are not presently available for rigorous two-center evaluation.

In the present work, we extend our approach to the case of one-electron heteronuclear quasimolecules: Bi--Au, U--Pb, and Cf--U.
The ground-state energy is evaluated in a wide range of internuclear distances, up to 1000 fm, in both two-center and monopole potentials.
Moreover, we consider three different monopole potentials, depending on the placement of the c.s.o.
For the heavy quasimolecules under consideration, the QED effects also play an important role.
We consider the leading self-energy and vacuum polarization contributions within the monopole approximation.

The relativistic units, $\hbar = c = m = 1$, and the Heaviside charge unit, $\alpha = e^2/(4\pi)$ (fine-structure constant), are used throughout the paper.

\section{Method}
We start with the Born--Oppenheimer approximation, in which the electron is described by the two-center Dirac equation,
\begin{gather}
    \label{eq:dirac_eq}
    \Big[ \vec{\alpha} \cdot \vec{p} + \beta + V(\vec{r}) \Big] \Psi_n(\vec{r}) = E_n \Psi_n(\vec{r})
\,, \\
    V(\vec{r}) = V^A_{\text{nucl}}(|\vec{r}-\vec{R}_A|) + V^B_{\text{nucl}}(|\vec{r}-\vec{R}_B|)
\,,
\end{gather}
where $\vec{r}$ and $\vec{R}_{A, B}$ are the position vectors of the electron and the nuclei, respectively; $V^{A, B}_{\text{nucl}}(r)$ are the spherically symmetric binding potentials generated by the nuclei; $\vec{p}$ is the momentum operator; $\vec{\alpha}$ and $\beta$ are the standard $4 \times 4$ Dirac matrices. 
The distance between the nuclei is denoted by $D = |\vec{R}_{A}-\vec{R}_{B}|$.
In this work, we use the Fermi model of the nuclear-charge distribution. 
The corresponding explicit formulas are well-known and can be found, e.g., in Ref.~{\cite{Parpia:1992}}.

The two-center (TC) potential is axially symmetric with respect to the internuclear axis.
In the spherical coordinate system $(r, \theta, \varphi)$ with the polar angle $\theta$ measured from this axis, the potential can be expanded into the following series:
\begin{equation}
    V(r, \theta) = \sum_{l} V_l(r) P_l(\cos\theta); \quad V_l(r) = \frac{2l + 1}{2} \int\limits_0^{\pi} V(r, \theta) P_l(\cos\theta) \sin\theta d\theta.
\end{equation}
\noindent The first term in this series, $V_0(r)$, corresponds to the widely used monopole approximation (MA).
Within this approximation, the initial axially symmetric problem is reduced to the spherically symmetric one.
Numerous methods developed for the atomic problem can be applied to solve the corresponding Dirac equation.
We use the dual-kinetically balanced finite-basis-set approach for both the TC and MA potentials; see Refs.~\cite{Shabaev:2004, Rozenbaum:2014, Kotov:2020, Kotov:2021} for more~details.

The spherical coordinates are used with three different placements of the c.s.o., namely: (1) in the middle between the nuclei, (2) in the center of the heavy nucleus, and (3) in the center of the light nucleus.
Whereas the TC approach provides the same results within numerical error bars, the MA values for the three different c.s.o. denoted by MA(1), MA(2), and MA(3), respectively, differ significantly. At large distances, the TC and different MA values often diverge qualitatively, while at $D \to 0$ they formally tend to the same limit.

In addition to the Dirac energies, we also evaluate the leading QED corrections---self-energy and vacuum polarization.
These terms are only treated within the monopole approximation; that is, the MA(1) potential is used in this case.
The computations follow the procedures discussed, e.g., in Refs.~\cite{Yerokhin:1999:800, glazov:2006:330, Artemyev:2007:173004, volotka:2008:062507, Artemyev:2013, malyshev:2022:012806}.
They are based on the expansion of the electron propagator in powers of the binding potential in order to isolate ultraviolet divergences and perform renormalization.

\section{Results}
\begin{figure}[H]
    \hspace{-15pt}\includegraphics[width=\textwidth]{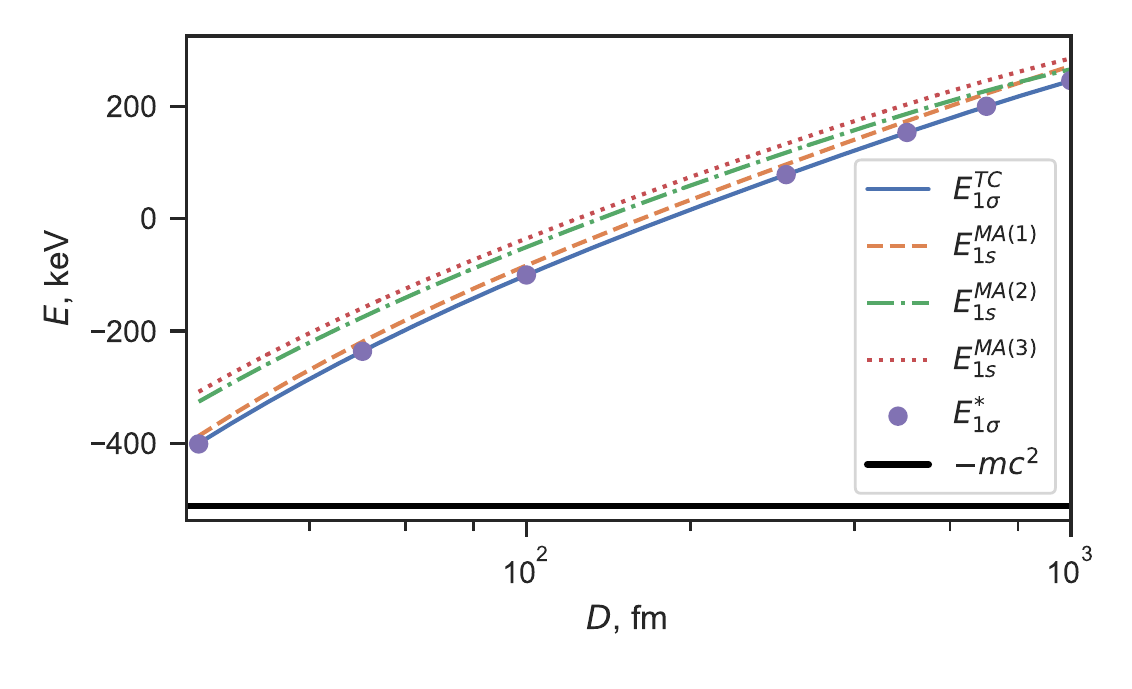}
    \caption{The ground-state Dirac energy of the U--Pb quasimolecule evaluated with the TC, MA(1), MA(2), and MA(3) potentials. $E^{*}_{1\sigma}$ corresponds to the data from Ref.~{\cite{Artemyev:2022}}}.\label{fig:energies}
\end{figure}

In Figure~\ref{fig:energies}, the ground-state energies of the Cf--U quasimolecule evaluated with the TC, MA(1), MA(2), and MA(3) potentials are presented.
Even though one may expect that all three monopole approximations converge at small internuclear distances, the obtained results show that the MA(1) energies are in fact much closer to the TC ones.
Nevertheless, the deviation between the TC and MA(1) grows towards the larger internuclear distances.
Furthermore, we note an almost constant difference between the MA(2) and the MA(3) results within the presented range.

There are two main sources for the total uncertainty of the obtained results: (1) the numerical error of the computational scheme, which is determined by the quality of the finite basis set employed in the practical calculations (basis-set error), and (2) the error associated with the uncertainties of the nuclear model and the root-mean-square radii (nuclear error).
The nuclear error provides the main contribution to the total uncertainty at the small internuclear distances and rapidly decreases towards the larger $D$.
Meanwhile, the basis-set error is rather small, and its value, e.g., for the U--Pb quasimolecule, does not exceed $10$~eV in the entire studied range.
Therefore, the total uncertainty at the small $D$ (up to $300$~fm) is almost completely determined by the nuclear error, while at the larger $D$ (from $500$ to $1000$~fm), it is determined by the basis-set error.

\begin{figure}[H]
     \hspace{-15pt}\includegraphics[width=\textwidth]{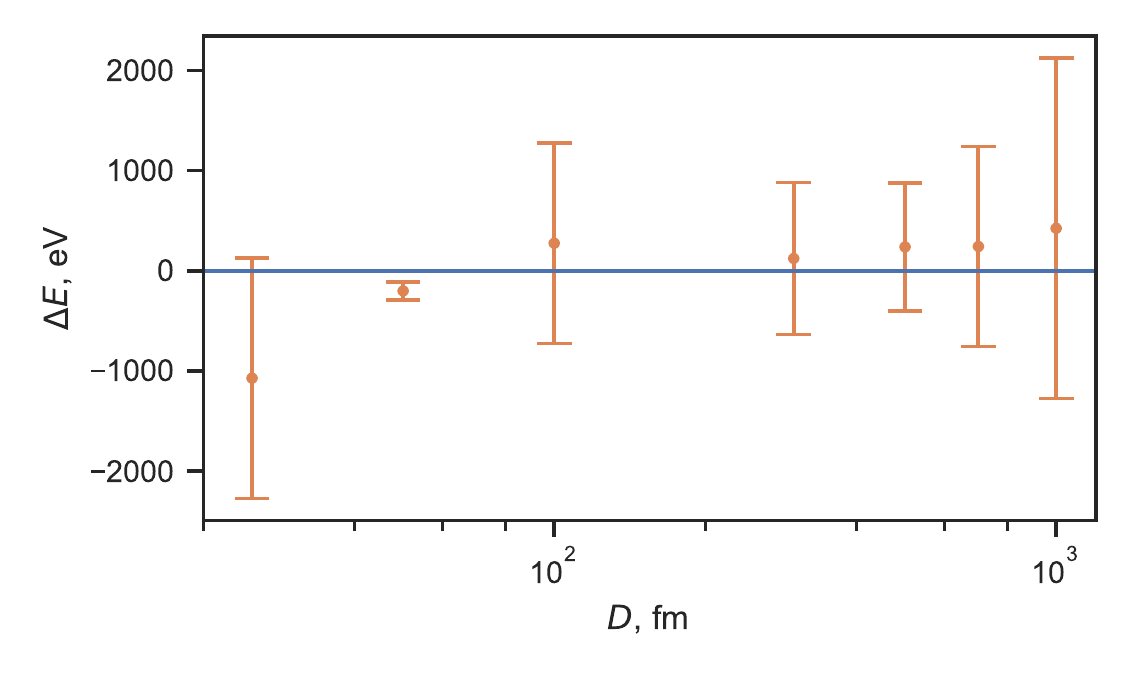}
    \caption{The ground-state binding energies (in eV) for the U--Pb quasimolecule from Ref.~\cite{Artemyev:2022} relative to the results obtained in the present work. See the text for details.\label{fig:errors}}
\end{figure}

In Figure~\ref{fig:errors}, we compare the ground-state energies for the U--Pb quasimolecule evaluated using the TC approach with the available data~\cite{Artemyev:2022}. 
The difference $\Delta E = E^{\text{Ref.\cite{Artemyev:2022}}}_{1\sigma} - E^{TC}_{1\sigma}$ is plotted.
All the values are in good agreement, except for the one with $D = 50$~fm.
For this internuclear distance, we estimate our total numerical error to be $\pm 30$~eV, which is three times smaller than the corresponding uncertainty presented in Ref.~\cite{Artemyev:2022}.
The reasons for this deviation are unclear to us.

The numerical data, including the self-energy and vacuum polarization contributions of all the quasimolecules under consideration, can be found in Table~\ref{tab:quasimolecule}.
We note that the difference between TC and MA(1) for the binding energies is significantly larger, more than an order of magnitude in most cases, than the total QED correction.
Thus, an evaluation of the Dirac energy within the rigorous two-center approach is crucial for the accurate determination of the electronic spectra.

\begin{table}[H]
    \caption{The ground-state Dirac energy, self-energy, and vacuum polarization contributions (in eV) for the one-electron Bi--Au, U--Pb, and Cf--U quasimolecules at different internuclear distances.\label{tab:quasimolecule}}
    \begin{tabularx}{\textwidth}{lRRRRRRR}
        \toprule
        {\boldmath{$D$}, \textbf{fm}} & {\boldmath{$E^{\text{\textbf{TC}}}_{1\sigma}$}} & {\boldmath{$E^{\text{\textbf{MA(1)}}}_{1s}$}} & {\boldmath{$E^{\text{\textbf{MA(2)}}}_{1s}$}} & {\boldmath{$E^{\text{\textbf{MA(3)}}}_{1s}$}} & {\boldmath{$\textbf{SE}^{\textbf{MA(1)}}_{1s}$}} & {\boldmath{$\textbf{VP}^{\textbf{MA(1)}}_{1s}$}} \\
        \midrule
        \multicolumn{7}{c}{Bi--Au}                                                                                                                                                                        \\
        \midrule
        15        & $-$237,546                   & $-$234,358                 & $-$196,293                 & $-$192,512                 & 6900                             & $-$6025                          \\
        25        & $-$171,018                   & $-$164,778                 & $-$128,136                 & $-$123,948                 & 5236                             & $-$3830                          \\
        50        & $-$79,797                    & $-$70,861                  & $-$39,310                  & $-$34,999                  & 3376                             & $-$1849                          \\
        100       & 6199                        & 16,418                     & 44,424                     & 48,863                     & 2071                             & $-$827                           \\
        300       & 137,579                      & 150,276                    & 173,231                    & 178,322                    & 800                              & $-$187                           \\
        500       & 198,231                      & 213,083                    & 230,800                    & 236,409                    & 451                              & $-$81                            \\
        700       & 237,280                      & 254,069                    & 266,215                    & 272,260                    & 290                              & $-$43                            \\
        1000      & 276,703                      & 296,127                    & 300,037                    & 306,662                    & 169                              & $-$21                            \\
        \midrule
        \multicolumn{7}{c}{U--Pb}                                                                                                                                                                         \\
        \midrule
        25        & $-$399,528                   & $-$386,390                 & $-$325,462                 & $-$307,855                 & 7378                             & $-$6230                          \\
        50        & $-$235,560                   & $-$218,680                 & $-$175,040                 & $-$158,648                 & 4446                             & $-$2722                          \\
        100       & $-$100,275                   & $-$83,130                  & $-$50,315                  & $-$35,068                  & 2570                             & $-$1113                          \\
        300       & 78,437                       & 96,786                     & 117,862                    & 133,471                    & 928                              & $-$227                           \\
        500       & 153,372                      & 173,899                    & 186,844                    & 203,557                    & 512                              & $-$95                            \\
        700       & 199,857                      & 222,639                    & 227,866                    & 245,689                    & 325                              & $-$50                            \\
        1000      & 245,476                      & 271,658                    & 265,930                    & 285,332                    & 188                              & $-$24                            \\
        \midrule
        \multicolumn{7}{c}{Cf--U}                                                                                                                                                                         \\
        \midrule
        50        & $-$491,640                   & $-$459,027                 & $-$383,320                 & $-$366,680                 & 6003                             & $-$4189                          \\
        80        & $-$329,591                   & $-$298,922                 & $-$242,676                 & $-$228,241                 & 4007                             & $-$2162                          \\
        100       & $-$263,819                   & $-$234,379                 & $-$184,810                 & $-$171,155                 & 3285                             & $-$1566                          \\
        200       & $-$91,283                    & $-$64,537                  & $-$30,027                  & $-$17,722                  & 1701                             & $-$551                           \\
        250       & $-$43,071                    & $-$16,584                  & 13,759                     & 25,933                     & 1349                             & $-$386                           \\
        500       & 91,335                       & 119,363                    & 134,508                    & 147,097                    & 594                              & $-$115                           \\
        700       & 149,097                      & 179,309                    & 184,223                    & 197,470                    & 372                              & $-$59                            \\
        1000      & 204,297                      & 238,194                    & 229,390                    & 243,724                    & 212                              & $-$28                            \\
        \bottomrule
    \end{tabularx}
\end{table}

\section{Conclusions}
In this work, the ground-state energies of the Bi--Au, U--Pb, and Cf--U quasimolecules at different internuclear distances, up to 1000 fm, were evaluated within the rigorous two-center approach.
The monopole approximation was also considered using three different placements of the coordinate system's origin: (1) in the middle between the nuclei, (2) in the center of the heavy nucleus, and (3) in the center of the light nucleus.
The results obtained within the two-center approach were found to be in good agreement with previous independent calculations for the Bi--Au and U--Pb quasimolecules.
The leading QED contributions, self-energy and vacuum polarization, were also evaluated within the monopole approximation.
Accurate theoretical predictions of the quasimolecular spectra require further development of the presented methods, including rigorous two-center evaluation of the QED contributions.

\acknowledgments
The one-electron energy calculations were funded by the Russian Science Foundation (Grant Number 21-42-04411). 
The studies on the QED corrections were supported by the Foundation for the Advancement of Theoretical Physics and Mathematics ``BASIS''.

\bibliographystyle{apsrev4-2}
\bibliography{references}

\end{document}